\documentstyle[preprint,pre,aps,epsfig]{revtex}

\begin{document}

\preprint{MA/UC3M/05/97}

\draft

\tighten

\title{dc motion of ac driven sine-Gordon solitons}

\author{Niurka R.\ Quintero$^{\dag}$ and Angel 
S\'anchez$^{\ddag}$}

\address{Grupo Interdisciplinar de Sistemas
Complicados,
Departamento de Matem\'aticas, 
\\ c/ Butarque 15,
Universidad Carlos III de Madrid, E-28911 Legan\'{e}s, Madrid, Spain}

\date{\today}

\maketitle

\begin{abstract}

We investigate the possibility of dc soliton motion sustained by pure ac 
driving in the sine-Gordon model. We show by means of the collective 
coordinate formalism that ac driving induces a net dc velocity whose 
modulus and direction 
depend on the driving phase. Numerical simulations of the full 
sine-Gordon equation confirm the correctness and accuracy of this prediction. 
Non trivial
cases when dc soliton motion is transformed into oscillatory as well
as the effects of damping are analyzed.
Our results settle a long standing issue
about the existence and characteristics of this phenomenon,
whose possible appearance in other systems is also discussed.

\end{abstract}


\section{Introduction}

It is well recognized that the study of nonlinear equations and their 
solutions is of great importance in many areas of physics. Among these,
the sine-Gordon equation has attracted the interest from mathematicians
and physicists due to not only its complete integrability but also its 
ubiquity as a model of nonlinear physical phenomena. Indeed, the sine-Gordon
equation is known to be a canonical model for a wide variety
of physical systems when topological solitons are present,
such as motion of dislocations in crystals \cite{Nabarro},
charge density waves \cite{Gruner}, 
solitons in magnetically ordered systems \cite{Kosevich},
epitaxial growth of thin films \cite{CW},
fluxon dynamics in long Josephson junctions \cite{Barone},
or DNA promoter dynamics \cite{DNA}.

One still unsolved issue regarding the application of external forces on
systems represented by the sine-Gordon model regards
the effects of pure ac driving and the possibility that it induces dc motion
of solitons. To clarify this question, 
in this Communication we analyze the corresponding perturbed sine-Gordon
equation
\begin{equation}
\begin{array}{l}
\phi_{tt} - \phi_{xx} + \sin(\phi) =
-\beta \phi_{t} + \epsilon f(t),
\end{array}
\label{ecua1}
\end{equation}
where subindices indicate derivative with respect to the corresponding 
variable, $\beta\phi_t$ is the usual damping term, and 
$f(t)=\sin(\delta t + \delta_{o})$ is an external periodic force describing,
for example, a long Josephson junction under the application of a uniform 
microwave field \cite{Matteo1}. 
In the following, we will only consider ac driving
(with or without dissipation)
acting on a single soliton. We note 
that the influence of an ac force like the one we 
study here acting on breather solutions of the sine-Gordon equation was
the subject of several papers by the end of the eighties, mostly 
from the view point of the use of a suitable frequency 
to stabilize breathers
in the presence of dissipation \cite{lomsam,kiv}.
The related problem of an 
ac driving in the presence of an additional dc driving has been already
considered and it is quite well understood
(see, e.g., \cite{Matteo1,Matteo2,Mario1} and references therein), 
whereas the effects of pure dc driving were established in the seminal
work of McLaughlin and Scott \cite{McL}. The issue of pure ac driving
acting on solitons, however,
has proven itself much more confusing, and a clear picture of soliton
behavior under this kind of force was still lacking. 

To our knowledge,
ac driven sine-Gordon solitons were 
first studied by Bonilla and Malomed \cite{BM} (and
subsequently in \cite{Malomed} for the Toda lattice with similar results),
who claimed that ac driving could support dc motion of solitons in a 
discrete (in space) sine-Gordon model, dissipation being crucial for 
this phenomenon. Unfortunately, as was shown in \cite{siete}, this result
was incorrect because of two reasons: First, it was obtained by means 
of a necessary, but not sufficient, existence condition, and second, the
authors used simultaneously and carelessly 
both the discrete and continuum limits of the model. Numerical simulations
confirmed that the dc motion predicted in \cite{BM} did not take place 
\cite{siete}. Recently, another study found a parameter region for the 
discrete sine-Gordon model (i.e., the Frenkel-Kontorova chain) where 
ac driving can induce dc motion. Such behavior can only occur
in the discrete model because its two ingredients arise from discreteness:
the Peierls-Nabarro barrier and its associated frequency \cite{nuevo}.
Hence this is a much more complicated (indeed, chaotic)
process involving attractor competition
and has nothing to do with the proposal of \cite{BM}. 

In this work,
we show that dc motion of solitons induced by pure ac driving in the 
continuum sine-Gordon equation is indeed 
possible: In the next Section,
we use a collective coordinate approach to compute
the soliton motion, finding that its mean velocity is a function of the
phase of the driving. We verify our results by means of numerical 
simulation, the comparison turning out to be satisfactory with an almost
perfect agreement. The final Section contains a discussion of our results
and comments on related subjects and generalizations to other systems.

\section{Collective coordinate analysis}

In order to perturbatively study eq.\ (\ref{ecua1}) we will resort to the usual
collective coordinate approach \cite{kiv2,yo}:
If $\beta$ and $\epsilon$ are small, we 
can approximate the one-soliton solution of eq.\ (\ref{ecua1}) by that of 
the unperturbed sine-Gordon equation ($\beta=\epsilon=0$) with time 
dependent parameters, i.e., we will make the following ansatz:
\begin{equation}
\begin{array}{l}
\displaystyle \phi(x,t) = 4 \>{\rm arctan}\left( \exp\left[\pm \frac{x-x_{o}(t)-X_{o}(t)}{\sqrt{1
-u^{2}(t)}}\right]\right), \\
\\
\displaystyle X_{o}(t) = \int_{0}^{t} {u(t')} dt',
\end{array}
\label{ecua2}
\end{equation}
where the positive (negative) sign corresponds to a kink (anti-kink) solution. 
We note that $X(t)=x_o(t)+X_{o}(t)$ [$u(t)$] 
has the meaning of the position (velocity) of the center of the
soliton, hence the name ``collective coordinate,'' and that the main assumption
underlying this approximation is that radiation effects induced by the 
perturbation are neglected. This will be verified {\em a posteriori} by 
comparing with the numerical simulations.

We now apply the method of McLaughlin and Scott \cite{McL} to obtain the 
equations of motion for the soliton center. The method, which they dubbed
energetic analysis, is a rather simple one and amounts to compute the 
variation of the energy and momentum
of the unperturbed sine-Gordon system due to the
perturbation and the same quantity for a soliton solution, subsequently 
imposing compatibility. This is already a classic procedure and details 
can be found in \cite{McL,kiv2,yo}. The resulting equations of motion 
are (again, the $\pm$ sign corresponds to a kink or an anti-kink)
\begin{equation}
\begin{array}{l}
\hspace{-.5cm} \displaystyle \frac{du}{dt} = -\frac{1}{4}
(1-u^{2}) (\pm \pi \sqrt{1-u^{2}} \epsilon f(t) + 4 \beta u ), \\
\\
\hspace{-.5cm}\displaystyle \frac{dX}{dt} =  u(t). \\
\\
\end{array}
\label{ecua3}
\end{equation}
We begin by discussing the dissipation-free case, i.e., we set $\beta=0$
in eqs.\ (\ref{ecua3}). In this case, the first equation of (\ref{ecua3})
can be exactly solved, yielding
\begin{equation}
\begin{array}{l}
u(t) = F[u(0),\delta_0]/\left(1+F[u(0),\delta_0]^2\right)^{1/2}, \\ \\
F[u(0),\delta_0]\equiv \displaystyle{\frac{u(0)}{\sqrt{1-u^{2}(0)}}} \pm
\displaystyle{\frac{\pi \epsilon}{4 \delta}}
[\cos(\delta t +\delta_{o})-\cos(\delta_{o})].
\end{array}
\label{ecua4}
\end{equation}
Equation (\ref{ecua4}) should now be integrated to obtain the final result,
namely the soliton center motion. This cannot be done in general; however,
when
$|u(0)| << 1$ and $|\frac{\pi \epsilon}{4 \delta}| << 1$, a good approximation
to the solution is given by 
\begin{equation}
\begin{array}{l}
\displaystyle X(t) = X(0) + [u(0) \mp \displaystyle{\frac{\pi \epsilon}{4 \delta}} \, \cos\, (\delta_{o})]\>t\pm \\ \\
\phantom{Hazme sitio}\pm \displaystyle{\frac{\pi \epsilon [\sin(\delta t + \delta_{o})-\sin(\delta_{o})]}{4 \delta^{2
}}}.
\end{array}
\label{ecua5}
\end{equation}
{}From eq.\ (\ref{ecua5}) it is evident that kinks and anti-kinks will move 
with net velocity $\{u(0) \mp [\pi \epsilon\cos\, (\delta_{o})]/
4\delta\}$ in a straight line, to which oscillations of frequency 
$\delta$ are overimposed. Therefore, in this approximation, an ac driving 
of frequency $\delta$ will {\em almost always} induce a dc motion of 
sine-Gordon solitons; indeed, only if 
\begin{equation}
\begin{array}{l}
\displaystyle u(0) = \pm \frac{\pi \epsilon\cos(\delta_{o})}{4 \delta}
\end{array}
\label{ecua6}
\end{equation}
solitons will remain oscillating around their initial position. It is 
important to note that this condition depends on the initial velocity 
of the soliton, $u(0)$; if $u(0)=0$, i.e., the soliton is initially at
rest, the condition is simply $\delta_0=(2n+1)\pi/2$, $n=0,\pm 1,\ldots$
But even if the soliton is initially moving, a phase can be chosen such 
that this dc motion is stopped and transformed to an oscillation, a non-trivial
and unexpected result.

We note that the above predictions have been obtained for the case of 
a slowly moving soliton and a small driving strength. It is possible,
however, to establish this result in general by means of an alternative
method based on the study of the Hamiltonian. This calculation is somewhat
more complicated, so we will report on in elsewhere \cite{segundo} and 
here we will just quote the final result, namely condition (\ref{ecua6})
for oscillatory motion becomes
\begin{equation}
\begin{array}{l}
\displaystyle \frac{u(0)}{\sqrt{1-u(0)^{2}}} =\pm \frac{\pi \epsilon \cos(\delta_{o})}
{4 \delta},
\end{array}
\label{ecua12}
\end{equation}
i.e., the only change with respect to the approximate condition previously
derived is the (otherwise expected) appearance of the Lorentz factor 
$\gamma={\sqrt{1-u(0)^{2}}}$. 

To conclude our analytical calculations for this problem, 
we now turn to the dissipative case.
Then, eq.\ (3) cannot be solved, not even for $u$. A slow
motion approximation ($|u(t)|<<1$) yields
\begin{equation}
\begin{array}{l}
\displaystyle u(t) = c \, \exp(-\beta t) \mp \\  \\
\phantom{haz sitio}\mp\displaystyle{\frac{\pi \epsilon}{4 (\beta^{2} + \delta^{2
})}}
[\beta \sin(\delta t + \delta_{o}) - \delta \cos(\delta t + \delta_{o})],
\end{array}
\label{ecua14}
\end{equation}
with 
\begin{equation}
\displaystyle c=u(0) \pm \frac{\epsilon \pi}{4 (\beta^{2} + \delta^{2})} [\beta \sin(\delta_{o})-
\delta \cos(\delta_{o})].
\label{ecua14bis}
\end{equation}
{}From the above, rather involved expression [or from 
the even more cumbersome one
for $X(t)$ which can be obtained by integrating (\ref{ecua14})], the main conclusion we can 
draw is that (within the small velocity approximation) solitons will never
exhibit dc motion except for a transient, after which they will reach a 
final oscillatory state around a point depending on the initial conditions.

\section{Numerical results}

\begin{figure}
\epsfig{file=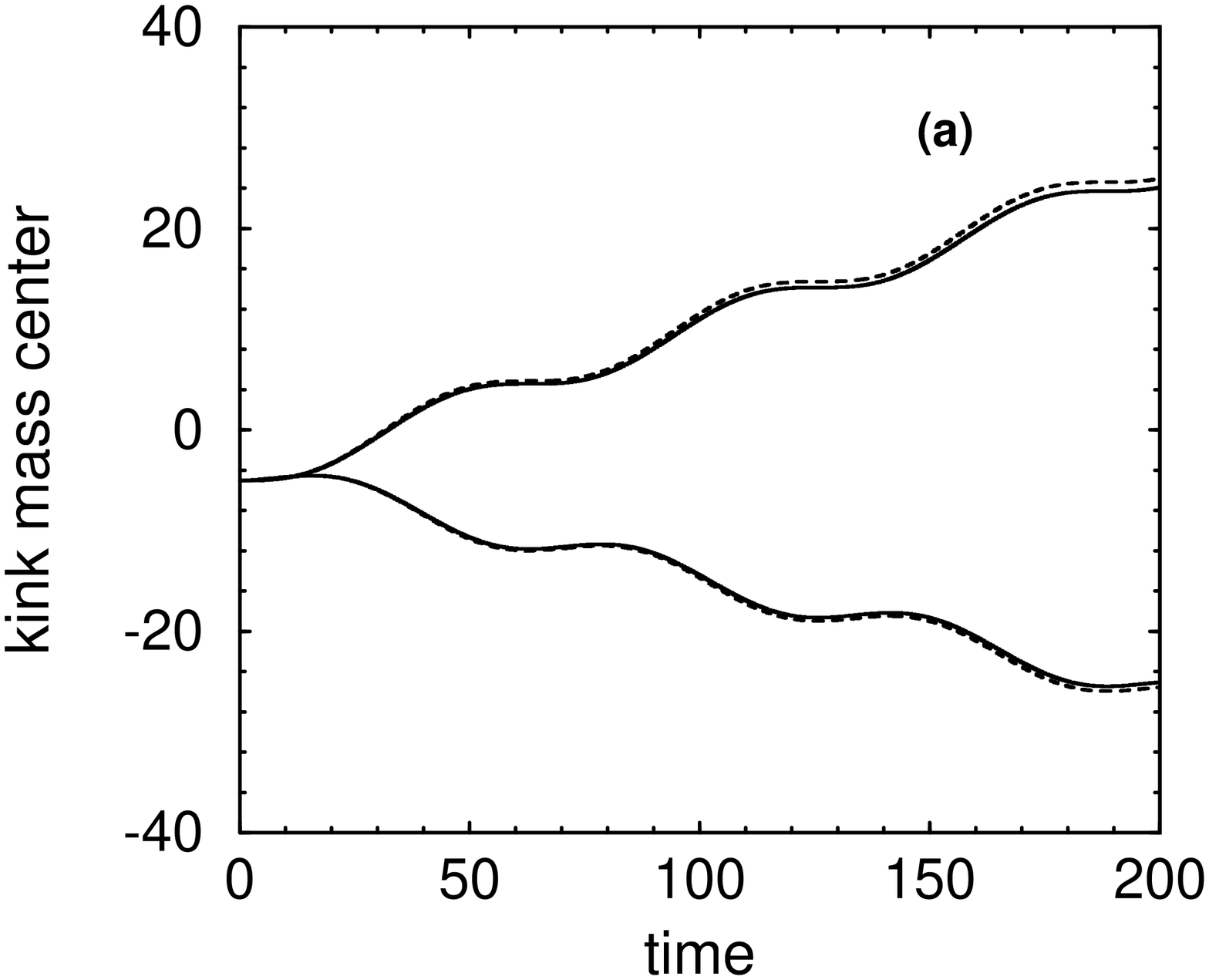, width=2.2in, angle=-90}
\epsfig{file=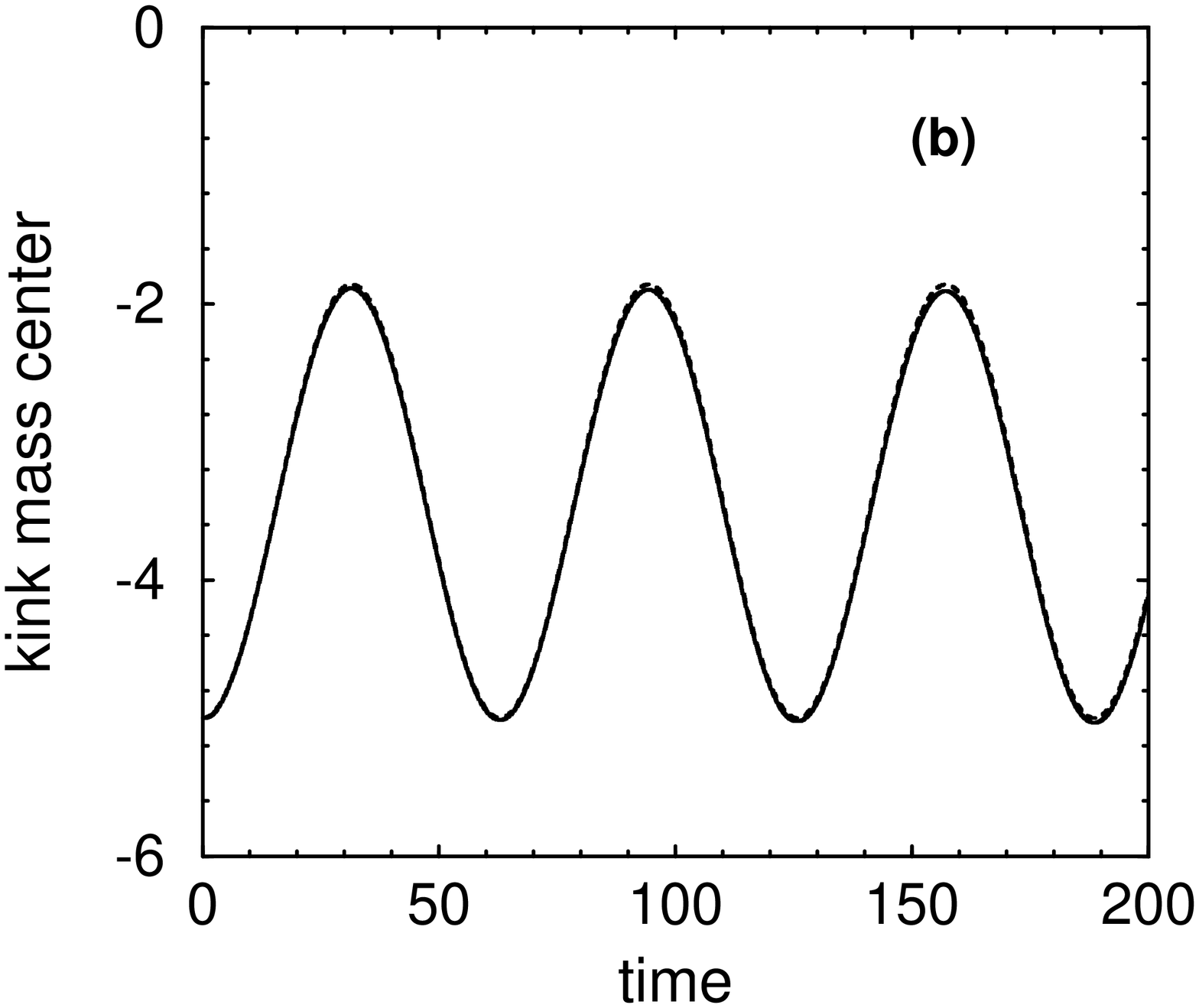, width=2.2in, angle=-90}
\caption[]{Verification of collective coordinate predictions in the
absence of dissipation. Simulation 
starts from a static kink located at 
$X(0)=-5$, and subject to an ac force given by $-0.02
\sin(0.1 t + \delta_{0})$. (a) $\delta_{0}=0$ (upper curve),
$\delta_{0}=\frac{3 \pi}{4}$ (lower curve); notice that 
the direction of motion is opposite in both cases.
(b) $\delta_{0}=\frac{\pi}{2}$, critical value exhibiting oscillatory motion.
Both in (a) and (b),
solid line corresponds to numerical integration of eq.\ (\ref{ecua1}),
dashed line is the analytical prediction (\ref{ecua5}).}
\label{graf2}
\end{figure}

All the results and predictions in the previous section involve some 
approximation, beginning with 
the collective coordinate hypothesis of negligible
radiation. Therefore, they are meaningless unless verified by numerical 
simulations of the full perturbed sine-Gordon problem (\ref{ecua1}). 
Simulations were carried out by means of a standard fourth order Runge-Kutta
algorithm \cite{nrecipies} with initial conditions given by an unperturbed
sine-Gordon soliton, at rest or with velocity $u(0)$, and boundary conditions
$\phi_x(L=\pm 50,t) = 0$.
The conclusion of our numerical simulation program was that the predictions
of the collective coordinate theory are very well verified by the full 
perturbed sine-Gordon equation. An example of the accuracy of our analytical
results is shown in Fig.\ \ref{graf2}, where we plot the results of simulations
in the absence of dissipation
for different values of the phase $\delta_0$ and a soliton (actually, a kink)
initially at rest. The high degree of agreement between theory and simulations
is apparent, confirming that dc motion of ac driven kinks is indeed the 
usual behavior, except for very particular choices of the phase of the 
driving. 

\begin{figure}
\epsfig{file=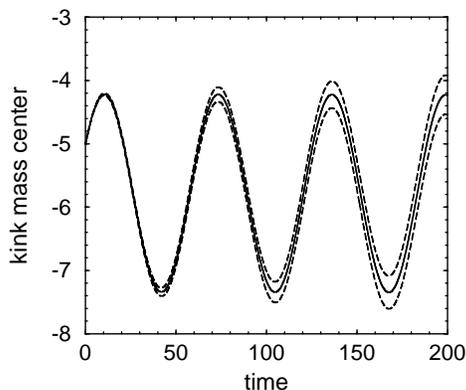, width=2.2in, angle=-90}
\caption{Stopping of a kink initially moving with velocity $u(0)=0.1335$ (solid
line) compared to the evolution of kinks moving with $u(0)=0.135$ (upper 
dashed curve) and $u(0)=0.132$ (lower dashed curve). The
ac driving had $\delta_0=\pi/6$; other parameters are $\delta=0.1$ and
$\epsilon=0.02$. The theoretically predicted 
critical velocity was $u(0)=0.1347$, and the corresponding theoretical 
evolution lies on top of the numerical curve for $u(0)=0.1335$.}
\label{graf3}
\end{figure}

The rest of the predictions are equally correct. Thus,
Fig.\ \ref{graf3} collects the outcome of simulations intended to
verify the accuracy of the prediction that solitons with specific velocities
can be stopped by a suitable choice of the driving phase. To verify this 
result, we proceed in the opposite way, i.e., we compute the critical velocity
for a given value of the phase, and then we obtain the same value numerically.
As can be seen from Fig.\ \ref{graf3}, the numerical value turns out to 
be $u(0)=0.1335$, to be compared to the predicted $u(0)=0.1347$, i.e.,
the accuracy is better than 1\%. Finally,
Fig.\ 3 is an example of the behavior of the system when dissipation
is present. We find numerically that, indeed, oscillatory motion is the only 
outcome of simulations for whatever value (not even small ones) of the 
initial velocity (which, in turn, agrees with the results in \cite{siete}). 

\section{Conclusions}

In summary, in this Communication we have analytically and numerically 
demonstrated the possibility of soliton dc motion induced by pure ac driving
in sine-Gordon systems. The direction and modulus of the corresponding 
velocity have been shown to depend on the driving phase, which for a
specific value (depending on the initial data) leads to pure ac motion 
even if the soliton was moving. The analytical results, obtained by means
of a collective coordinate approach, agree with the 
numerical ones within an accuracy better than 1\%. We have also found that
this effect is only possible in the absence of dissipation, and that the 
final steady state for all damping values is the oscillatory one. This 
conclusion finally clarifies the problem of ac driving effects on 
solitons and opens the way to its applications in different 
technological contexts, such as, e.g., Josephson junctions 
\cite{Barone,Matteo1}.

To close this section, some comments are in order about the existence of 
this phenomenon in other systems. We believe that the features we have
found here are generic in the collective coordinate sense, i.e., for 
systems whose solitons exhibit a clear and robust particle-like behavior
(see, e.g., \cite{kiv2,yo} and references therein). However, care must be
taken when dealing with solitons possesing inner degrees of freedom, 
such as the sine-Gordon breather or the $\phi^4$ solitary wave, as 
the frequencies pertaining to these degrees of freedom will necessarily 
interact in a complicated manner with the driving frequency. It is clear
that this competition can give rise to chaotic phenomena, similar to 
that found in \cite{nuevo}, overruling the 
simple scenario we have depicted here. On the other hand, if those frequencies
are very different, our results should once again hold. Preliminary 
analytical and numerical investigations support this conclusion 
\cite{segundo}.

\section*{Acknowledgments}

We thank Fernando Falo for discussions
and for independent numerical checks of our predictions,
and Esteban Moro and Franz Mertens for conversations on this work.
Work at Legan\'es has been supported by CICyT (Spain) grant MAT95-0325.

\begin{figure}
\epsfig{file=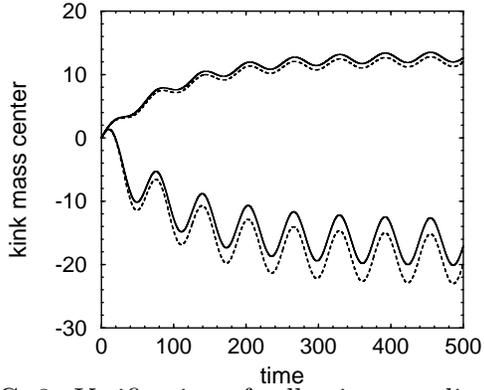, width=2.2in, angle=-90}
\caption{Verification of collective coordinate predictions with
dissipation. Shown are two examples of soliton motion subject to 
an ac force given by $\epsilon
\sin(0.1 t + \delta_{0})$, with $\epsilon=0.01$ (upper curve) and 
$\epsilon=0.05$ (lower curve); other parameters are $\beta=0.01$, 
$\delta_{0}=0.1$ and $u(0)=0.2$.
Solid lines correspond to numerical integration of eq.\ (\ref{ecua1}),
dashed lines are the analytical predictions obtained by 
integrating (\ref{ecua14}).}
\label{graff}
\end{figure}

\end{document}